\documentclass[aps,pra,twocolumn,superscriptaddress,showpacs,floatfix]{revtex4-2}

\usepackage{hyperref}
\usepackage{xcolor}
\usepackage{subfig}
\usepackage{natbib}
\hypersetup{
    colorlinks=true,    % Enables colored links
    linkcolor=blue,     % Sets the color of internal links (e.g., references)
    citecolor=blue,     % Sets the color of citation links
    filecolor=blue,     % Sets the color of file links
    urlcolor=blue       % Sets the color of URL links
}

\usepackage[normalem]{ulem}
\usepackage{physics}

\usepackage{float}
\usepackage{graphics}      % standard graphics specifications
\usepackage{graphicx}      % alternative graphics specifications
\usepackage{url}           % for on-line citations
\usepackage{bm}            % special 'bold-math' package

\usepackage{amsmath}
\usepackage{amssymb}
\usepackage{cleveref}
\usepackage{siunitx}

\newcommand{\half}{{1\over 2}}
\renewcommand{\phi}{\varphi}
\renewcommand{\epsilon}{\varepsilon}

\begin{document}
\title{Measuring the spin of a spin-1/2 and getting a result of 7}

\author{Joseph McGowan IV*} 
\affiliation{Department of Physics, University of Toronto, 60 St. George Street, Toronto ON, M5S 1A7, Canada}

\author{Nicholas Mantella*}
\affiliation{Department of Physics, University of Toronto, 60 St. George Street, Toronto ON, M5S 1A7, Canada}

\author{Noah Baker}
\affiliation{Department of Physics, University of Toronto, 60 St. George Street, Toronto ON, M5S 1A7, Canada}

\author{Aephraim M. Steinberg}
\affiliation{Department of Physics, University of Toronto, 60 St. George Street, Toronto ON, M5S 1A7, Canada}

\begin{abstract}
    The weak value of an operator is the average result of measuring that operator on a quantum system given specified initial and final states. The overlap between these states appears in the denominator of the weak value, and thus if that overlap is small, the weak value can be arbitrarily large. In 1988, Albert, Aharonov, and Vaidman proposed a weak Stern-Gerlach experiment which would be able to measure a so-called anomalous weak value, a measurement result which lies outside of the eigenvalue spectrum of the operator being measured. Though anomalous weak values have been measured in numerous optical systems, here we present the first such realization of this experiment as originally proposed. By placing a Bose-Einstein condensate of \(^{87}\)Rb in a magnetic gradient and performing an unlikely postselection, we measure an effective magnification of a weak magnetic momentum kick (\(\sim 8\) \si{\micro\meter}/s) by a factor of 14. We also demonstrate that increasing the momentum kick decreases the ``weakness'' of the measurement and reduces the amplification.
\end{abstract}

\maketitle

\section{Introduction}

Measurement has been at the heart of quantum mechanics since its inception a century ago. Rules for how the 
measurement process affects a quantum system, and for how results are obtained,
are simply postulated in the modern quantum formalism. In the 1930s, von Neumann was the first to consider a 
quantum picture of the measurement device by considering it as a separate quantum system interacting with 
the system to be measured \cite{VNmeas}. This measurement device could be imagined to be in some localized state whose 
width is much smaller than the separation between the eigenvalues of the system observable \(\hat{A}\) being measured. 
In this scenario, obtaining a measurement result of some eigenvalue \(a\) corresponds to finding the state of the
measurement device displaced by an amount proportional to \(a\).

This model for quantum measurement, now known as a von Neumann measurement, writes the interaction with the 
Hamiltonian
\begin{equation}
\hat{H}_I = g \hat{A}_S \otimes \hat{p}_P,
\end{equation}
where \(\hat{A}\) is the system operator to be measured, \(\hat{p}_P\) is the momentum operator for the measurement
device (usually called the pointer), and the parameter
\(g\) controls the strength of the coupling. Clearly, if the  
initial state is an eigenstate of \(\hat{A}\) with eigenvalue \(a\), 
the resulting pointer shift \(\pi\), the unitary acting on the 
pointer is proportional to \(e^{a \hat{p}_P}\), which simply displaces the pointer by \(a\). 
In 1988, Albert, Aharonov and Vaidman examined a different limit of this picture, where the pointer state is 
instead wide compared to the spread of possible shifts created by the eigenvalue spectrum of \(\hat{A}\)
% , as depicted in \cref{fig:cartoon} 
\cite{AAV}. They showed that, 
if the system starts in an
initial state \(\ket{i}\) and is projected onto a final state \(\ket{f}\) following this ``weak'' measurement of 
\(\hat{A}\), the average pointer position \(\pi\) is instead displaced by \(g \expval{A}_w\), where
\begin{equation}
\expval{A}_w = {\bra{f}\hat{A}\ket{i}\over \braket{f}{i}}
\end{equation}
is called the weak value and is well-defined when \(\ket{i}\) and \(\ket{f}\) are not orthogonal. 
% The previous statement is true when \(g\) is a constant, but in the case where \(g = g(t)\), a function
% with support inside of some defined interval representing the start and end of the measurement, the pointer shift
% remains proportional to the weak value. 
Since the pointer position is highly uncertain in this limit, a single measurement will not, in general, give the weak 
value; however, one can repeat many trials and obtain the weak value to the desired level of precision.

If the overlap of the initial and final states is small, but the matrix element of \(\hat{A}\) is large, this 
seemingly predicts a measurement result which can be arbitrarily large, and specifically outside the eigenvalue 
range of \(\hat{A}\). Indeed, when \(\ket{i}\) and \(\ket{f}\) are nearly orthogonal, the weak value 
diverges.
It was shown the following year that, though \(\pi\) cannot in general be arbitrarily large
given a particular pointer uncertainty, anomalous values may still be measurable. In particular,
Duck and Sudarshan showed that a superposition of Gaussians 
(or similar functions) whose peaks all lie within the eigenvalue range of \(\hat{A}\) can nevertheless interfere 
to give a Gaussian of the same width whose peak lies far outside that eigenvalue range \cite{Duck}. 
% It is at first counter-intuitive that two curves, whose peaks define the
% endpoints of 
% an interval, may be combined to yield another curve whose peak lies outside of that interval. If they are 
% added with real coefficients, this is indeed impossible; however, allowing the coefficients to be complex, 
% as is the case with interfering quantum states, permits this to happen \cite{Duck}.

This effect, known as weak value amplification (WVA) or anomalous weak values, was demonstrated soon after its 
introduction in 1991 by Ritchie, Story, and Hulet \cite{hulet}. In the years since, it has been demonstrated 
in numerous other photonic systems \cite{parabox, polarizationwva, deflectionwva, matin}, 
with superconducting qubits \cite{pastfuture, SCQWVA}, 
and using neutron interferometry \cite{neutrons, neutronstrong}, the
last marking the first observation of the effect using massive particles.
The effect has been shown to provide a 
metrological advantage, amplifying a small signal to be measured, in systems with certain types of noise or 
limitations \cite{noise, wvatechnoise, pangnoise, wvasaturation}. There are proposals to use WVA for 
measurements of neutron electric dipole moments \cite{edm} or for an enhanced version of LIGO \cite{ligo}.
However, despite the wide range of experiments investigating and measuring weak values, 
the experiment originally proposed by Aharonov et al. to illustrate the possibility of amplification, that
of weakly measuring a component of a particle's spin, has not been realized. This is, in large part, related to
the relative difficulty in observing interference with matter compared to light, since WVA is fundamentally
an interference effect. With laser light, it has been possible for decades to obtain coherence lengths at 
easily accessible length scales on the order of centimeters or meters. Modern lasers exceed this by many orders
of magnitude. Matter waves, on the other hand, generally maintain coherence over much shorter length scales.
For example, a typical BEC of alkali atoms near its transition temperature (\(\sim\) 100 nK) has a coherence length of 
less than 1 \si{\micro\meter}.
The work in this article therefore represents the closest realization of the weak Stern-Gerlach experiment 
as originally proposed, using a 
BEC of \(^{87}\)Rb. This is the first anomalous measurement of a particle's spin with a true continuous 
probe, as well as the first anomalous weak value observed using atoms.

\section{Theory}
\label{wsgtheory}

% A standard Von Neumann measurement considers an interaction between two quantum systems, referred to as the system,
% to be measured, and the pointer, which conveys the measurement result. The interaction Hamiltonian is given by 
% \[\hat{H}_I = g \hat{A}_S \otimes \hat{p}_P,\]
% where \(\hat{A}\) is the system operator to be measured, \(\hat{p}_P\) is the pointer momentum, and the parameter
% \(g\) controls the strength of the coupling. The conceptual framework of this model can be seen by considering an 
% initial state which is an eigenvalue of \(\hat{A}\); the resulting pointer shift \(\pi\), following a strong 
% measurement, will simply be proportional to the corresponding eigenvalue. 
% AAV showed that, in the limit of a weak measurement, the average 
% measurement result for initial state \(\ket{i}\) and final state \(\ket{f}\) is given by the weak value
% \[\expval{\hat{A}}_w = {\bra{f}\hat{A}\ket{i}\over \braket{f}{i}}.\]
In a Stern-Gerlach experiment, the Hamiltonian represents the interaction of a particle's spin with a 
magnetic field, given by \(\vb{\mu}\cdot\vb{B}\). In particular, for a field gradient in the \(z\) direction, the 
Hamiltonian takes the form \(\hat{\mu}_z \partial_z B \hat{z} = g \hat{\sigma}_z \hat{z}\), where \(\hat{\sigma}_z\)
is the Pauli operator, and the factor \(g\) includes the gradient strength and the effective interaction time. 
The system under study is the particle's spin, and the pointer ``position'', the degree of freedom shifted by 
coupling to the spin, is the particle's momentum. 

Ordinarily, the gradient strength or interaction time is increased sufficiently so that the momentum shift between
two spin components is large compared to the momentum width of the atomic beam or cloud. If the initial state is a 
superposition between multiple spin states, this results in spots which
are well-separated in position following some subsequent free evolution, and therefore well-distinguished measurement
results.
For a weak Stern-Gerlach experiment, one can imagine reducing the gradient strength or interaction time such that the 
spots are no-longer well-separated. The amplitudes of the two spin components are then able to interfere before the 
final strong postselection. 

We consider an initial system state of \(\ket{i} = \ket{+x} \propto \ket{+z} + \ket{-z}\) and a final system 
state \(\ket{f(\delta)} = 
\hat{R}_y(\delta)\ket{-x}\); that is, a state which 
has been rotated by an angle \(\delta\) away from \(\ket{-x}\). For \(\delta = 0\), therefore, 
the initial and final states are orthogonal.
For small \(\delta\), \(\braket{f(\delta)}{i}\sim \delta \ll 1\). However, \(\hat{\sigma}_z\ket{+x} = \ket{-x}\), 
and so 
the weak value \(\bra{f(\delta)}\hat{\sigma}_z\ket{i}/\braket{f(\delta)}{i} \sim 1/\delta\), allowing for an 
anomalously large 
pointer shift. For the pointer, the initial position state (i.e. pointer momentum state) is a Thomas-Fermi profile 
given by
\[
TF(z) \equiv
\begin{cases}
    \sqrt{1-\left({x\over \sigma}\right)^2} & |x| < \sigma\\
    0 & \text{otherwise}
\end{cases}
\]
This is chosen so that the initial momentum (i.e. pointer position) is 0.
In the case of a spin-1/2 particle, the two spin
states confer equal and opposite momentum shifts \(\pm \delta p\) 
on the corresponding components of the pointer state, resulting in the intermediate state \((\hbar=1)\)
\begin{equation} 
\label{eq:intermediatestate}
    \ket{\psi} = {1\over\sqrt{2}}\left(e^{iz\delta p/}TF(z)\ket{+x} + 
e^{-iz\delta p}TF(z)\ket{-x}.\right)
\end{equation}
The final postselected state is obtained by projecting the spin portion of this state onto \(\ket{f(\delta)}\),
and the momentum expectation value of the resulting state gives the pointer shift \(\pi\) on the ``position'' of
the pointer. Since the state lives in a tensor product space \(\mathcal{H}_\text{position}\otimes 
\mathcal{H}_\text{spin}\), this means that the pointer shift \(\pi\) is obtained from 
the expectation value of the operator 
\begin{equation}
\label{eq:otimesp}
    p_f = p \otimes \ketbra{f(\delta)}. 
\end{equation}

\begin{figure}
    \centering

    \subfloat{
        \includegraphics[width=0.9\linewidth]{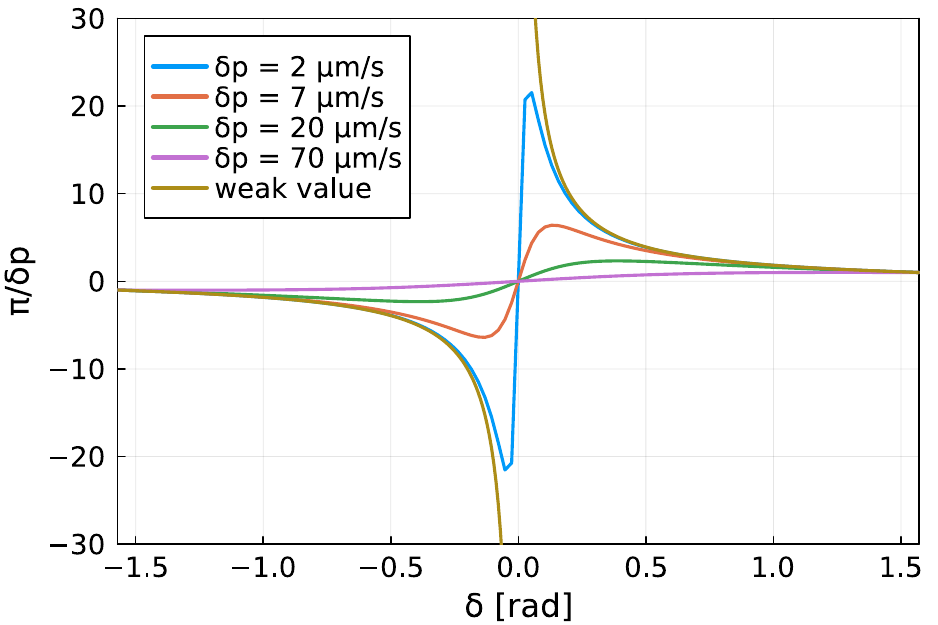}}

    \subfloat{
        \includegraphics[width=0.9\linewidth]{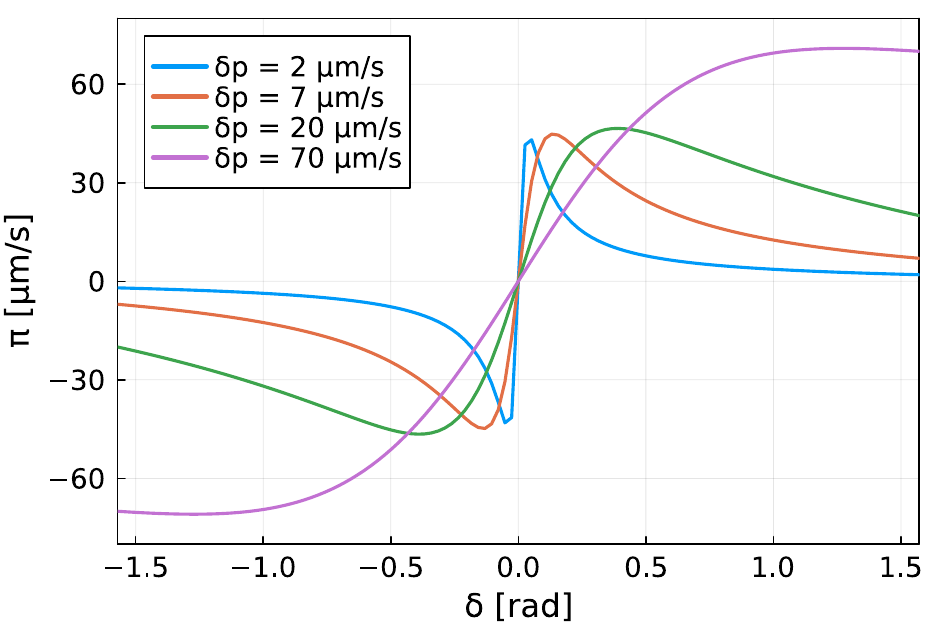}}
    \caption{Pointer shifts and amplification factors for various
    values of the momentum kick \(\delta p\). Above is the bare pointer shift, i.e. 
    \(\expval{p_f}_{\ket{\psi}} = \pi_f\), for the
    postselected state, and below is \(\pi_f/\delta p\). Shifts were 
    calculated using \(TF(z)\) with \(\sigma = 25\) \si{\micro\meter}. The lower plot also shows the weak value 
    \(\bra{f(\delta)}\sigma_z\ket{i}/\braket{f(\delta)}{i}\).}
    \label{fig:delpshifts}
\end{figure}

\Cref{fig:delpshifts} shows simulations of \(\pi\) for various applied momentum shifts 
\(\delta p\), along with the calculated weak value. For this measurement, the pointer shift is given by 
\(\delta p \expval{A}_w\).
When \(\delta p\) is large, there is no 
amplification for any \(\delta\); \(\pi\) is bounded by \(\pm \delta p\), and spin rotations 
merely swap the states. Decreasing \(\delta p\) causes the magnitude of the pointer shift to 
exceed \(\delta p\) in both directions. The maximum shift, however, is always bounded for small \(\delta p\); 
the maximum amplified shift depends weakly on \(\delta p\) but is of order \(1/\sigma\).

It should be noted that \(\delta p\) is ``large'' or ``small'' in comparison not to \(\Delta p\) of the initial 
wavepacket but instead to \(\hbar/\Delta x\). This is analogous to the coherence length, and may be thought of as 
a ``coherence momentum'': the scale over which different momentum components can interfere. For a Heisenberg-limited
state, these two scales coincide, but otherwise the coherence momentum is generally much smaller than \(\Delta p\).
It is by using a BEC that we are able to reach a regime where the coherence momentum is large enough to create 
observable interference.

The amplification factor \(\pi/\delta p\) shows a feature that
grows larger and narrower as the measurement becomes weaker, getting closer to the weak value prediction; however,
though the weak value diverges at \(\delta = 0\), the amplification factor instead passes through zero. This 
is due to the breakdown of the weak approximation, as discussed in \cite{Duck}. The width of the region
where the breakdown occurs can be arbitrarily small as \(\delta p \rightarrow 0\) but never zero. The 
shift \(\pi\) always goes to \(\pm \delta p\) and the amplification factor to \(\pm 1\) on the edges of the plot,
as the pointer is simply measuring the applied shift with no postselection, having at most swapped the states
\(\ket{+z}\) and \(\ket{-z}\), equivalent to flipping the sign of \(\delta p\). 
Since the 
measurement results \(\pi = \pm \delta p\) correspond to \(S_z \sim \pm \half\), an amplification factor which is
outside \([1,-1]\) effectively corresponds to measuring a value for \(S_z\) which is outside its eigenvalue range.

Depending on the initial shape of the cloud, the shift \(\pi\) can also be calculated analytically. If the 
initial state is a Gaussian with width \(\sigma\), then the expectation value of momentum for the 
shifted state can be evaluated using elementary integral identities. One finds that (noting \(\hbar =1\))
\begin{equation}
    \expval{p} = {\delta p \cos(\delta/2)\sin(\delta/2) \over 1 - e^{-\delta p^2 \sigma^2/4}\cos(\delta)}.
    \label{eq:pEV}
\end{equation}
Differentiating this, one finds that the pointer shift reaches a maximum value of 
\begin{equation}
    \pi_\text{max} = {\delta p \over 2 \sqrt{1 - e^{\delta p^2 \sigma^2/2}}}.
    \label{eq:pmax}
\end{equation}
For large \(\delta p\), meaning \(\delta p\sigma \gg 1\), this result is simply equal to \(\delta p\), reflecting
the fact that a large momentum shift represents a strong measurement. No amplification can be observed since 
the pointer shift does not exceed the applied momentum shift. For small \(\delta p\), this expression becomes 
independent of \(\delta p\) and is equal to \(1/\sigma \sqrt{2}\). As \(\delta p\) is reduced, the maximum 
pointer shift is not, and the amplification factor grows. 

\section{Methods}

We begin with a BEC of about 5-10 \(\cross 10^3\) atoms of \(^{87}\)Rb in a crossed optical dipole trap. The trap
is composed of one beam, called the waveguide, which creates a weak harmonic trap in the longitudinal direction
(\(\omega = 3.7\) Hz) and tight confinement in the radial direction (\(\omega \approx 300\) Hz), and another
beam which provides confinement along the waveguide's longitudinal axis. The energy
associated with the repulsive interactions of the atoms causes the cloud's initial size to change with the
atom number. The initial cloud size \(\Delta x\) in the longitudinal direction ranges from 18 to 33 
\si{\micro\meter}.

The condensate is formed in the \(\ket{F=2,m_F=2}\) state. Following condensation, the atoms are transferred
to the state \(\ket{2,0}\) using two-photon RF adiabatic rapid passage (ARP). Since this transfer is not 
perfectly efficient, we use a shelving technique to ensure that there are no stray atoms in \(\ket{2,1}\) or 
\(\ket{2,2}\). First, we perform a microwave \(\pi\) pulse on the \(\ket{2,0} \rightarrow 
\ket{1,0}\) transition. Then, we illuminate the atoms for 0.5 ms with light resonant on the \(F=2
\rightarrow F'=3\) D2 transition. Any population in the \(F=2\) manifold, left over from imperfect ARP or 
\(\pi\) pulses, will scatter many photons and therefore escape from the trap; this allows us to create a 
pure initial state
for our condensate with no contributions from any other \(m\) levels by performing another \(\pi\) pulse. 

To create the initial state for the weak measurement, we perform a \(\pi/2\) pulse to create an 
equal superposition of \(\ket{2,0}\)
and \(\ket{1,0}\). In order to perform the weak 
Stern-Gerlach, we must weakly couple their spin and momentum. As the two states are both \(m=0\) states,
to first order a magnetic field gradient will not create any shift in their relative momentum. However, their
energies also depend on the field to second order, given by the well-known Breit-Rabi formula:
\[E(B) = - {\Delta E_0\over 2(2I+1)} + g_I \mu_B m B \pm {\Delta E_0 \over 2}\sqrt{
1 + {4mx \over 2I+1} + x^2}. \]
Here \(\Delta E_0 = 6.835\) GHz is the zero-field hyperfine splitting, \(I = 3/2\) is the nuclear spin, \(g_I\) and
\(g_J\) are the relevant g-factors, \(\mu_B\) the Bohr magneton, and \(x = (g_J - g_I)\mu_B B/\Delta E\)
\cite{steck}. The \(\pm\) is + for \(F=2\) and - for \(F=1\). For
the states in question, \(m = 0\), so the linear dependence vanishes. However, differentiating the resulting
formula with respect to the field magnitude shows that \(\ket{2,0}\) and \(\ket{1,0}\) will experience equal
and opposite forces in a magnetic gradient. However, unlike the force on other \(m\) states,
the force, at first order, depends not only on the gradient strength but also the local field strength. 
For a gradient in the z direction,

\begin{equation}
    \pdv{E(B)}{z}\Bigg|_\text{m=0} = \pdv{|\vb{B}|}{z}\:{x^2 \Delta E \over 2|B|\sqrt{1+x^2}}.
\end{equation}
If \(\vb{B} \neq B_z \hat{z}\) and there is some other field \(B_\perp\) in the xy plane, then 
\[\pdv{|\vb{B}|}{z} = {B_z\over \sqrt{B_z^2 + B_\perp^2}}\pdv{B_z}{z}.\]
In effect, the field gradient provides a weak coupling between the total spin \(\vb{F}\) and the atomic
momentum. However, since the momentum shifts of the two states are equal and opposite, the coupling is 
functionally identical to that of a spin-\(\half\) in a standard Stern-Gerlach experiment.

The Breit-Rabi formula can also be used to calculate the transition frequency \(\Delta E(B)\) between the 
m=0 states. This provides a reliable and simple way of calibrating the field strengths \(B_z\) and \(B_\perp\).
Using Ramsey spectroscopy, we can locate the transition frequency to a precision on the order of 10 Hz, 
corresponding to a field uncertainty at the mG level. 

After applying the gradient pulse, the state of the condensate will be approximately given by
\begin{equation}
\ket{\psi} = {1\over \sqrt{2}}\left(e^{i\delta p z}\ket{2,0} + e^{-i\delta p z}\ket{1,0}\right),
\end{equation}
where \(\delta p\) is the momentum shift imparted by the gradient. In a standard
Stern-Gerlach experiment, the atomic cloud or beam would now be allowed to expand, and after a sufficiently
long time, position would become correlated with momentum. However, due to the long times of flight which
would be necessary, the atomic cloud would expand significantly in this time. At low atom numbers, this 
causes a reduction in density which in turn increases error in imaging metrics. Instead, we release the atoms
into the waveguide, which provides a weak harmonic potential. Since harmonic oscillators simply rotate states 
in phase space, after 1/4 of a period in this harmonic oscillator, the momentum shift above is
mapped onto a position shift between the two hyperfine components. 

In order to observe amplification, we must post-select on a state which has small overlap with the initial 
superposition \(\ket{i} = (\ket{2,0} + \ket{1,0})/\sqrt{2}\). Immediately after the gradient pulse, we 
perform an additional microwave pulse of variable duration on the transition between the m=0 states. Since 
the imaging light is only absorbed by atoms in the \(F=2\) manifold, this variable pulse serves to map a chosen
final state \(\bra{f(\delta)}\) onto the imaging state \(\bra{2,0}\). 

In addition to imaging the final state \(\ket{f(\delta)}\), we also image the orthogonal state \(\ket{n(\delta)}\). 
After the
final microwave pulse, this state is mapped onto \(\ket{1,0}\), and so any population in this state is unaffected 
while the state \(\ket{2,0}\) is imaged. To image this state, we first perform ARP to transfer the population
 to \(\ket{2,0}\) and then perform 
a second, identical imaging pulse. Imaging both states allows us to measure the difference in the centers of 
the two clouds. As the atoms are released into the waveguide, they may gain a non-zero momentum, which appears
as common-mode noise in the centers of the two clouds after 1/4 period. Measuring the difference significantly
reduces the effect of these fluctuations on our statistics. This also allows us to compute the post-selection
probability and hence the overlap \(\braket{f(\delta)}{i}\) for each experimental shot. The ARP transfer results in an
additional expansion time of 5.6 ms between the two images, which can be accounted for. 

\section{Results}

% \begin{figure}
%     \centering

%     \begin{subfigure}[b]{0.45\textwidth}
%         \includesvg[width=0.9\linewidth]{wsgdata_wv_lines.svg}
%     \end{subfigure}

%     \begin{subfigure}[b]{0.45\textwidth}
%         \includesvg[width=0.9\linewidth]{wsgdata_alg.svg}
%     \end{subfigure}
%     \caption{Experimental data showing the measured momentum shift (pointer position 
%     shift) between the two spin components of the BEC. At each value of \(\delta\), each pair of images for a 
%     single experimental shot is fitted and the centers subtracted, then this value for all pairs is averaged 
%     together. Errorbars represent the standard error on the mean for 60 pairs each. Above, the data 
%     are shown with the weak value predictions based on the unamplified kick. Below, the same data are 
%     compared with analytic predictions for a Gaussian initial state of equivalent size.}
%     \label{fig:wsgdata}
% \end{figure}

\begin{figure}
    \centering
    \includegraphics[width=0.9\linewidth]{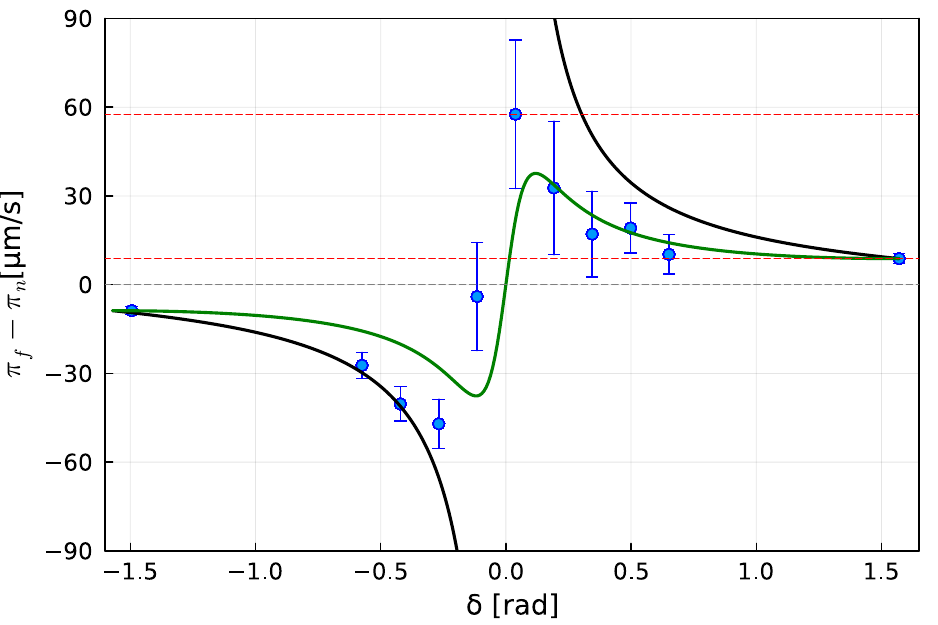}
    \caption{Experimental data showing the measured momentum shift (pointer position 
    shift) between the two spin components of the BEC. At each value of \(\delta\), each pair of images for a 
    single experimental shot is fitted and the centers subtracted, then this value for all pairs is averaged 
    together. Errorbars represent the standard error on the mean for 60 pairs each. The data 
    are shown with the weak value predictions based on the unamplified kick (in black) and with analytic 
    predictions for a Gaussian initial state of equivalent size (in green).}
    \label{fig:wsgdata}
\end{figure}

\Cref{fig:wsgdata} shows experimental data for a cloud with an initial Thomas-Fermi radius of 
31 \si{\micro\meter}, along with 
the corresponding weak value prediction as well as the analytic prediction given by \cref{eq:pmax}.
The data show reasonable agreement with the 
%simulations,
weak value, and the amplification of the pointer shift is evident. Each absorption image is fitted using a 2D 
Thomas-Fermi profile. The centers from corresponding image pairs are subtracted, and the resulting 
differences are averaged for each value of \(\delta\). This method gives error bars (standard error) which are 
5-10 times smaller than subtracting the averages of the first images and second images. Due to the low 
post-selection probabilities near \(\delta = 0\), the atomic density can be so low that it is not possible to 
meaningfully fit atomic profiles in some images. To find and eliminate such images, we find the extreme edges of 
high-density clouds for \(\delta = \pm \pi/2\) and omit any images whose fits give centers outside of this 
range. This is justified since the density of the postselected cloud is everywhere less than the density 
of the interfering clouds and must therefore be zero outside of this range. However, the presence of such 
points illustrates the difficulty in fitting the atomic profiles at such low densities. When the atom number
drops too low, the shot noise due to dark counts on the camera sensor pollutes the image and makes it 
harder to accurately find the center of the atomic cloud. Fluctuations in the overall atom number may 
therefore occasionally drop the density below the point where the center can reliably be found. 
The presence of such spurious points effectively 
broadens the distribution and therefore increases the error in the mean.

The key features of the weak value predictions can nevertheless be clearly seen. At the left and right edges, for 
\(|\delta| \approx \pi/2\), the pointer shift is simply \(\pm \delta p\) for the two atomic states. The 
difference between the centers is therefore \(\pm 2\delta p\). However, near \(\delta = 0\), the overlap
\(\braket{f}{i}\) becomes small, and the weak value grows; this breaks down right at \(\delta=0\), as 
mentioned in \cref{wsgtheory}. The measured momentum shifts show excellent agreement with the 
expected shape, well exceeding \(2 \delta p\) in magnitude on either side of \(\delta = 0\) and passing
through 0 at \(\delta = 0\). 
% At its largest, the pointer shift is 7 \(\pm\) 3 times larger than at, 
% representing an anomalous measurement result for \(\sigma_z\) of 3.5 \(\pm 1.5\); that is, the 
% measured pointer shift is 7 times larger than the applied momentum kick. 
The point with the largest pointer shift \(\pi_f - \pi_n\), 
at \(\delta = 0.38\), has a shift of 14(6) \(\delta p\) from 0, 
7(3) times larger than the shift observed at \(\delta = \pm \pi/2\). These two points are indicated in the left 
side of \cref{fig:wsgdata} with horizontal green lines. Here \(\delta p = 4.4(5)\) 
\si{\micro\meter}/s. As \(\pi_n \approx 0\) near \(\delta = 0\),
this corresponds to a weak value of 7(3) for \(\sigma_z\), as though a measurement of the z component of this 
effective spin-\(\half\) yielded a result of \(m_z=7\).
By performing 
the unlikely postselection, the remaining atoms experienced a coupling to the field gradient which was 
14 times stronger than in the absence of the postselection. 

The bottom of \cref{fig:wsgdata} shows the same data alongside the prediction of \cref{eq:pEV}. 
Though the scale of amplification seen in the data seems to slightly exceed that in the theory, the width of the 
region in which the weak approximation breaks down agrees well with the prediction. The slight mismatch may 
simply be due to the fact that our atomic cloud is not a Gaussian; the comparison is done by setting the 
Gaussian width \(\sigma\) in \cref{eq:pEV} equal to \(\Delta x = \sqrt{\expval{x^2}}\) for a Thomas-Fermi
wavefunction with the measured width.

\begin{figure}
    \centering
    \includegraphics[width=0.9\linewidth]{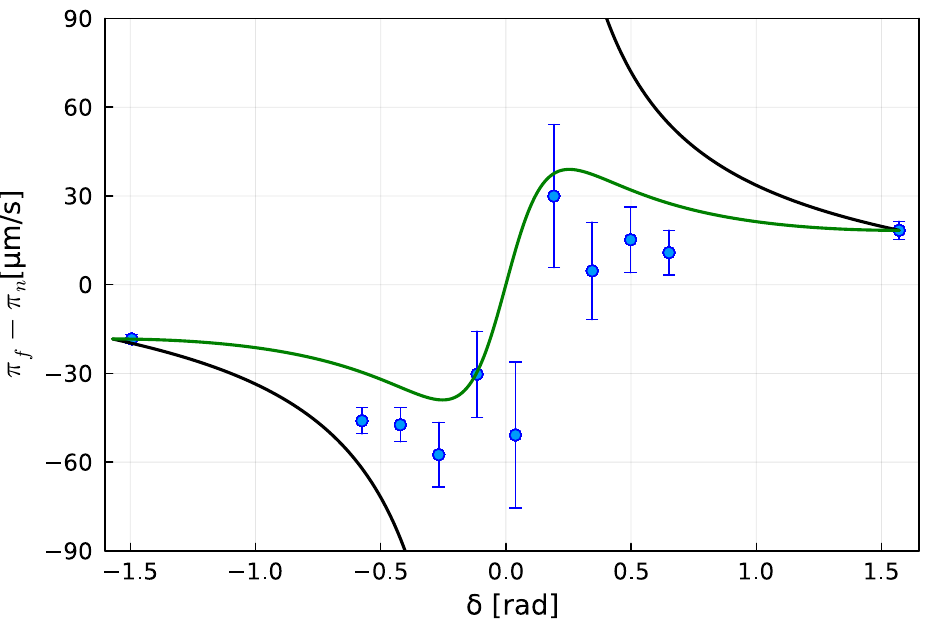}
    \caption{Experimental data for a cloud with a similar initial 
    size and atom number to the data in \cref{fig:wsgdata} but with a momentum kick approximately 2 times 
    larger (\(\delta p = 9.2 \pm 0.8\) \si{\micro\meter\per\second}). 
    The observed amplification is correspondingly smaller, and the deviation from the weak value prediction
    (shown in black) occurs sooner. The green curve is calculated from \cref{eq:pEV} as in \cref{fig:wsgdata}.}
    \label{fig:wsgdata2}
\end{figure}

% [other figure] shows similar results with a larger initial size and with a larger applied momentum 
% kick \(\delta p\), respectively. For a larger initial cloud, the corresponding pointer uncertainty 
% is smaller, and so a given momentum kick would make the measurement  ``less weak'' and therefore result 
% in a smaller amplification factor. A larger momentum kick has a similar effect; however, since the 
% breakdown point of the weak approximation near \(\delta = 0\) is determined by the pointer width, only
% the former case should increase the width of the overall feature. The measured pointer shifts suggest 
% that these predictions are correct; however, the large error bars currently preclude quantitative 
% comparisons. The difficulty in precisely determining these momentum shifts largely come from the low
% atomic density, as mentioned above. A future experiment might leverage a Feshbach resonance to decrease
% the atomic interaction energy prior to the momentum kick; this would result in a smaller cloud width
% following expansion, and therefore a higher atomic density for the same number of atoms. 

\Cref{fig:wsgdata2} shows similar results for a cloud with a larger momentum kick \(\delta p\). 
For a given initial cloud size, a larger momentum kick makes the measurement ``less weak''. The data correspondingly
deviate from the weak value prediction more strongly and at a larger range of values of \(\delta\), and the 
maximum amplification factor is also smaller. If instead the initial size is made smaller, this increases 
the momentum uncertainty, and therefore a given momentum kick would be made ``more weak''. The expectation 
in this case is that the data should first deviate from the weak value prediction at a smaller value of 
\(\delta\) and reach a larger amplification factor; the peaks should be higher and narrower. However, 
we were not able to observe this; the only available method to significantly reduce the cloud size required 
also reducing the atom number, resulting in a loss of signal.
A future experiment might leverage a Feshbach resonance to decrease
the atomic interaction energy prior to the momentum kick; this would result in a smaller cloud width
following expansion, and therefore a higher atomic density for a given number of atoms compared to this experiment.

% However, due to large interaction energy present in the cloud at the moment of the momentum kick, the 
% width of the atomic clouds in the final images is large compared to the position shift created by the 
% momentum kick, and so the errors remain somewhat large.

\section{Conclusion}

We have performed the first demonstration of weak value amplification using atoms, and the closest version 
of the experiment proposed alongside the original presentation of weak values. Using the second-order Zeeman
shift of the BEC in a magnetic gradient, we saw that the momentum shift of the condensate was increased, upon 
postselection, by a factor of 14 compared to the shift actually applied by the gradient. This effect was 
successfully able to ``measure'' a momentum shift at the \si{\micro\meter}/s scale, a rarely accessed velocity
regime. This experiment 
demonstrates that, despite their low coherence volumes, it is possible to observe anomalous weak values in 
atomic systems. This type of experiment has been called ``infeasible'' in the past \cite{neutrons}; 
we have shown that it is feasible.

Experiments like this one may be useful in the future for magnetic gradiometry applications. It may also be
useful in searches for small magnetic moments, where the amplification technique may be used to make the 
effect more visible. However, any such experiment would likely require some refinement of the experimental 
techniques used here. If the signal-to-noise ratio could be improved, the errors on the amplified points where
the postselection is unlikely would be more useful for metrology. This could be achieved in a subsequent experiment
using an atomic species with a convenient Feshbach resonance, which would enable a higher density for a given
atom number. 

\section{Acknowledgments}

The first two authors, marked with asterisks, contributed equally to this work. 
The authors would like to thank Robert Flack, Basil Hiley, and David Spierings for helpful discussions.

% \begin{figure}
%     \centering
%     \includegraphics[width=0.9\linewidth]{good data run.png}
%     \caption{Data for initial size of some microns. Simulations \ds{not on plot yet} show good agreement. Error 
%     bars on some points are smaller than the marker size.}
%     \label{fig:good data}
% \end{figure}

% \begin{figure}
%     \centering
%     \includegraphics[width=0.9\linewidth]{ok comparison.png}
%     \caption{Data for a push of 1.5 ms and 1.648 ms \ds{should we just round to 1.65? the 48 is just because
%     of the timing resolution of rampage. Though, will probably just convert this to momentum units anyway.}}
%     \label{fig:pushsize}
% \end{figure}

\begin{small}
% Bibliography:
\begingroup
\raggedright
\bibliographystyle{IEEEtran}
\bibliography{biblio}

\begin{thebibliography}{19}%
\makeatletter
\providecommand \@ifxundefined [1]{%
 \@ifx{#1\undefined}
}%
\providecommand \@ifnum [1]{%
 \ifnum #1\expandafter \@firstoftwo
 \else \expandafter \@secondoftwo
 \fi
}%
\providecommand \@ifx [1]{%
 \ifx #1\expandafter \@firstoftwo
 \else \expandafter \@secondoftwo
 \fi
}%
\providecommand \natexlab [1]{#1}%
\providecommand \enquote  [1]{``#1''}%
\providecommand \bibnamefont  [1]{#1}%
\providecommand \bibfnamefont [1]{#1}%
\providecommand \citenamefont [1]{#1}%
\providecommand \href@noop [0]{\@secondoftwo}%
\providecommand \href [0]{\begingroup \@sanitize@url \@href}%
\providecommand \@href[1]{\@@startlink{#1}\@@href}%
\providecommand \@@href[1]{\endgroup#1\@@endlink}%
\providecommand \@sanitize@url [0]{\catcode `\\12\catcode `\$12\catcode `\&12\catcode `\#12\catcode `\^12\catcode `\_12\catcode `\%12\relax}%
\providecommand \@@startlink[1]{}%
\providecommand \@@endlink[0]{}%
\providecommand \url  [0]{\begingroup\@sanitize@url \@url }%
\providecommand \@url [1]{\endgroup\@href {#1}{\urlprefix }}%
\providecommand \urlprefix  [0]{URL }%
\providecommand \Eprint [0]{\href }%
\providecommand \doibase [0]{https://doi.org/}%
\providecommand \selectlanguage [0]{\@gobble}%
\providecommand \bibinfo  [0]{\@secondoftwo}%
\providecommand \bibfield  [0]{\@secondoftwo}%
\providecommand \translation [1]{[#1]}%
\providecommand \BibitemOpen [0]{}%
\providecommand \bibitemStop [0]{}%
\providecommand \bibitemNoStop [0]{.\EOS\space}%
\providecommand \EOS [0]{\spacefactor3000\relax}%
\providecommand \BibitemShut  [1]{\csname bibitem#1\endcsname}%
\let\auto@bib@innerbib\@empty
%</preamble>
\bibitem [{\citenamefont {von Neumann}(1955)}]{VNmeas}%
  \BibitemOpen
  \bibfield  {author} {\bibinfo {author} {\bibfnamefont {J.}~\bibnamefont {von Neumann}},\ }\bibinfo {title} {Mathematical foundations of quantum mechanics}\ (\bibinfo  {publisher} {Princeton University Press},\ \bibinfo {year} {1955})\ pp.\ \bibinfo {pages} {271--283}\BibitemShut {NoStop}%
\bibitem [{\citenamefont {Aharonov}\ \emph {et~al.}(1988)\citenamefont {Aharonov}, \citenamefont {Albert},\ and\ \citenamefont {Vaidman}}]{AAV}%
  \BibitemOpen
  \bibfield  {author} {\bibinfo {author} {\bibfnamefont {Y.}~\bibnamefont {Aharonov}}, \bibinfo {author} {\bibfnamefont {D.~Z.}\ \bibnamefont {Albert}},\ and\ \bibinfo {author} {\bibfnamefont {L.}~\bibnamefont {Vaidman}},\ }\bibfield  {title} {\bibinfo {title} {How the result of a measurement of a component of the spin of a spin-1/2particle can turn out to be 100},\ }\href {https://doi.org/10.1103/physrevlett.60.1351} {\bibfield  {journal} {\bibinfo  {journal} {Physical Review Letters}\ }\textbf {\bibinfo {volume} {60}},\ \bibinfo {pages} {1351–1354} (\bibinfo {year} {1988})}\BibitemShut {NoStop}%
\bibitem [{\citenamefont {Duck}\ \emph {et~al.}(1989)\citenamefont {Duck}, \citenamefont {Stevenson},\ and\ \citenamefont {Sudarshan}}]{Duck}%
  \BibitemOpen
  \bibfield  {author} {\bibinfo {author} {\bibfnamefont {I.~M.}\ \bibnamefont {Duck}}, \bibinfo {author} {\bibfnamefont {P.~M.}\ \bibnamefont {Stevenson}},\ and\ \bibinfo {author} {\bibfnamefont {E.~C.~G.}\ \bibnamefont {Sudarshan}},\ }\bibfield  {title} {\bibinfo {title} {The sense in which a “weak measurement” of a spin-½ particle’s spin component yields a value 100},\ }\href {https://doi.org/10.1103/physrevd.40.2112} {\bibfield  {journal} {\bibinfo  {journal} {Physical Review D}\ }\textbf {\bibinfo {volume} {40}},\ \bibinfo {pages} {2112–2117} (\bibinfo {year} {1989})}\BibitemShut {NoStop}%
\bibitem [{\citenamefont {Ritchie}\ \emph {et~al.}(1991)\citenamefont {Ritchie}, \citenamefont {Story},\ and\ \citenamefont {Hulet}}]{hulet}%
  \BibitemOpen
  \bibfield  {author} {\bibinfo {author} {\bibfnamefont {N.~W.~M.}\ \bibnamefont {Ritchie}}, \bibinfo {author} {\bibfnamefont {J.~G.}\ \bibnamefont {Story}},\ and\ \bibinfo {author} {\bibfnamefont {R.~G.}\ \bibnamefont {Hulet}},\ }\bibfield  {title} {\bibinfo {title} {Realization of a measurement of a ‘“weak value”’},\ }\href {https://doi.org/10.1103/physrevlett.66.1107} {\bibfield  {journal} {\bibinfo  {journal} {Physical Review Letters}\ }\textbf {\bibinfo {volume} {66}},\ \bibinfo {pages} {1107–1110} (\bibinfo {year} {1991})}\BibitemShut {NoStop}%
\bibitem [{\citenamefont {Resch}\ \emph {et~al.}(2004)\citenamefont {Resch}, \citenamefont {Lundeen},\ and\ \citenamefont {Steinberg}}]{parabox}%
  \BibitemOpen
  \bibfield  {author} {\bibinfo {author} {\bibfnamefont {K.}~\bibnamefont {Resch}}, \bibinfo {author} {\bibfnamefont {J.}~\bibnamefont {Lundeen}},\ and\ \bibinfo {author} {\bibfnamefont {A.}~\bibnamefont {Steinberg}},\ }\bibfield  {title} {\bibinfo {title} {Experimental realization of the quantum box problem},\ }\href {https://doi.org/10.1016/j.physleta.2004.02.042} {\bibfield  {journal} {\bibinfo  {journal} {Physics Letters A}\ }\textbf {\bibinfo {volume} {324}},\ \bibinfo {pages} {125–131} (\bibinfo {year} {2004})}\BibitemShut {NoStop}%
\bibitem [{\citenamefont {Pryde}\ \emph {et~al.}(2005)\citenamefont {Pryde}, \citenamefont {O’Brien}, \citenamefont {White}, \citenamefont {Ralph},\ and\ \citenamefont {Wiseman}}]{polarizationwva}%
  \BibitemOpen
  \bibfield  {author} {\bibinfo {author} {\bibfnamefont {G.~J.}\ \bibnamefont {Pryde}}, \bibinfo {author} {\bibfnamefont {J.~L.}\ \bibnamefont {O’Brien}}, \bibinfo {author} {\bibfnamefont {A.~G.}\ \bibnamefont {White}}, \bibinfo {author} {\bibfnamefont {T.~C.}\ \bibnamefont {Ralph}},\ and\ \bibinfo {author} {\bibfnamefont {H.~M.}\ \bibnamefont {Wiseman}},\ }\bibfield  {title} {\bibinfo {title} {Measurement of quantum weak values of photon polarization},\ }\bibfield  {journal} {\bibinfo  {journal} {Physical Review Letters}\ }\textbf {\bibinfo {volume} {94}},\ \href {https://doi.org/10.1103/physrevlett.94.220405} {10.1103/physrevlett.94.220405} (\bibinfo {year} {2005})\BibitemShut {NoStop}%
\bibitem [{\citenamefont {Dixon}\ \emph {et~al.}(2009)\citenamefont {Dixon}, \citenamefont {Starling}, \citenamefont {Jordan},\ and\ \citenamefont {Howell}}]{deflectionwva}%
  \BibitemOpen
  \bibfield  {author} {\bibinfo {author} {\bibfnamefont {P.~B.}\ \bibnamefont {Dixon}}, \bibinfo {author} {\bibfnamefont {D.~J.}\ \bibnamefont {Starling}}, \bibinfo {author} {\bibfnamefont {A.~N.}\ \bibnamefont {Jordan}},\ and\ \bibinfo {author} {\bibfnamefont {J.~C.}\ \bibnamefont {Howell}},\ }\bibfield  {title} {\bibinfo {title} {Ultrasensitive beam deflection measurement via interferometric weak value amplification},\ }\bibfield  {journal} {\bibinfo  {journal} {Physical Review Letters}\ }\textbf {\bibinfo {volume} {102}},\ \href {https://doi.org/10.1103/physrevlett.102.173601} {10.1103/physrevlett.102.173601} (\bibinfo {year} {2009})\BibitemShut {NoStop}%
\bibitem [{\citenamefont {Hallaji}\ \emph {et~al.}(2016)\citenamefont {Hallaji}, \citenamefont {Feizpour}, \citenamefont {Dmochowski}, \citenamefont {Sinclair},\ and\ \citenamefont {Steinberg}}]{matin}%
  \BibitemOpen
  \bibfield  {author} {\bibinfo {author} {\bibfnamefont {M.}~\bibnamefont {Hallaji}}, \bibinfo {author} {\bibfnamefont {A.}~\bibnamefont {Feizpour}}, \bibinfo {author} {\bibfnamefont {G.}~\bibnamefont {Dmochowski}}, \bibinfo {author} {\bibfnamefont {J.}~\bibnamefont {Sinclair}},\ and\ \bibinfo {author} {\bibfnamefont {A.~M.}\ \bibnamefont {Steinberg}},\ }\href {https://doi.org/10.48550/ARXIV.1612.04920} {\bibinfo {title} {How the result of counting one photon can turn out to be a value of 8}} (\bibinfo {year} {2016})\BibitemShut {NoStop}%
\bibitem [{\citenamefont {Campagne-Ibarcq}\ \emph {et~al.}(2014)\citenamefont {Campagne-Ibarcq}, \citenamefont {Bretheau}, \citenamefont {Flurin}, \citenamefont {Auffèves}, \citenamefont {Mallet},\ and\ \citenamefont {Huard}}]{pastfuture}%
  \BibitemOpen
  \bibfield  {author} {\bibinfo {author} {\bibfnamefont {P.}~\bibnamefont {Campagne-Ibarcq}}, \bibinfo {author} {\bibfnamefont {L.}~\bibnamefont {Bretheau}}, \bibinfo {author} {\bibfnamefont {E.}~\bibnamefont {Flurin}}, \bibinfo {author} {\bibfnamefont {A.}~\bibnamefont {Auffèves}}, \bibinfo {author} {\bibfnamefont {F.}~\bibnamefont {Mallet}},\ and\ \bibinfo {author} {\bibfnamefont {B.}~\bibnamefont {Huard}},\ }\bibfield  {title} {\bibinfo {title} {Observing interferences between past and future quantum states in resonance fluorescence},\ }\bibfield  {journal} {\bibinfo  {journal} {Physical Review Letters}\ }\textbf {\bibinfo {volume} {112}},\ \href {https://doi.org/10.1103/physrevlett.112.180402} {10.1103/physrevlett.112.180402} (\bibinfo {year} {2014})\BibitemShut {NoStop}%
\bibitem [{\citenamefont {Groen}\ \emph {et~al.}(2013)\citenamefont {Groen}, \citenamefont {Ristè}, \citenamefont {Tornberg}, \citenamefont {Cramer}, \citenamefont {de~Groot}, \citenamefont {Picot}, \citenamefont {Johansson},\ and\ \citenamefont {DiCarlo}}]{SCQWVA}%
  \BibitemOpen
  \bibfield  {author} {\bibinfo {author} {\bibfnamefont {J.~P.}\ \bibnamefont {Groen}}, \bibinfo {author} {\bibfnamefont {D.}~\bibnamefont {Ristè}}, \bibinfo {author} {\bibfnamefont {L.}~\bibnamefont {Tornberg}}, \bibinfo {author} {\bibfnamefont {J.}~\bibnamefont {Cramer}}, \bibinfo {author} {\bibfnamefont {P.~C.}\ \bibnamefont {de~Groot}}, \bibinfo {author} {\bibfnamefont {T.}~\bibnamefont {Picot}}, \bibinfo {author} {\bibfnamefont {G.}~\bibnamefont {Johansson}},\ and\ \bibinfo {author} {\bibfnamefont {L.}~\bibnamefont {DiCarlo}},\ }\bibfield  {title} {\bibinfo {title} {Partial-measurement backaction and nonclassical weak values in a superconducting circuit},\ }\bibfield  {journal} {\bibinfo  {journal} {Physical Review Letters}\ }\textbf {\bibinfo {volume} {111}},\ \href {https://doi.org/10.1103/physrevlett.111.090506} {10.1103/physrevlett.111.090506} (\bibinfo {year} {2013})\BibitemShut {NoStop}%
\bibitem [{\citenamefont {Sponar}\ \emph {et~al.}(2015)\citenamefont {Sponar}, \citenamefont {Denkmayr}, \citenamefont {Geppert}, \citenamefont {Lemmel}, \citenamefont {Matzkin}, \citenamefont {Tollaksen},\ and\ \citenamefont {Hasegawa}}]{neutrons}%
  \BibitemOpen
  \bibfield  {author} {\bibinfo {author} {\bibfnamefont {S.}~\bibnamefont {Sponar}}, \bibinfo {author} {\bibfnamefont {T.}~\bibnamefont {Denkmayr}}, \bibinfo {author} {\bibfnamefont {H.}~\bibnamefont {Geppert}}, \bibinfo {author} {\bibfnamefont {H.}~\bibnamefont {Lemmel}}, \bibinfo {author} {\bibfnamefont {A.}~\bibnamefont {Matzkin}}, \bibinfo {author} {\bibfnamefont {J.}~\bibnamefont {Tollaksen}},\ and\ \bibinfo {author} {\bibfnamefont {Y.}~\bibnamefont {Hasegawa}},\ }\bibfield  {title} {\bibinfo {title} {Weak values obtained in matter-wave interferometry},\ }\bibfield  {journal} {\bibinfo  {journal} {Physical Review A}\ }\textbf {\bibinfo {volume} {92}},\ \href {https://doi.org/10.1103/physreva.92.062121} {10.1103/physreva.92.062121} (\bibinfo {year} {2015})\BibitemShut {NoStop}%
\bibitem [{\citenamefont {Denkmayr}\ \emph {et~al.}(2018)\citenamefont {Denkmayr}, \citenamefont {Dressel}, \citenamefont {Geppert-Kleinrath}, \citenamefont {Hasegawa},\ and\ \citenamefont {Sponar}}]{neutronstrong}%
  \BibitemOpen
  \bibfield  {author} {\bibinfo {author} {\bibfnamefont {T.}~\bibnamefont {Denkmayr}}, \bibinfo {author} {\bibfnamefont {J.}~\bibnamefont {Dressel}}, \bibinfo {author} {\bibfnamefont {H.}~\bibnamefont {Geppert-Kleinrath}}, \bibinfo {author} {\bibfnamefont {Y.}~\bibnamefont {Hasegawa}},\ and\ \bibinfo {author} {\bibfnamefont {S.}~\bibnamefont {Sponar}},\ }\bibfield  {title} {\bibinfo {title} {Weak values from strong interactions in neutron interferometry},\ }\href {https://doi.org/10.1016/j.physb.2018.04.014} {\bibfield  {journal} {\bibinfo  {journal} {Physica B: Condensed Matter}\ }\textbf {\bibinfo {volume} {551}},\ \bibinfo {pages} {339–346} (\bibinfo {year} {2018})}\BibitemShut {NoStop}%
\bibitem [{\citenamefont {Sinclair}\ \emph {et~al.}(2017)\citenamefont {Sinclair}, \citenamefont {Hallaji}, \citenamefont {Steinberg}, \citenamefont {Tollaksen},\ and\ \citenamefont {Jordan}}]{noise}%
  \BibitemOpen
  \bibfield  {author} {\bibinfo {author} {\bibfnamefont {J.}~\bibnamefont {Sinclair}}, \bibinfo {author} {\bibfnamefont {M.}~\bibnamefont {Hallaji}}, \bibinfo {author} {\bibfnamefont {A.~M.}\ \bibnamefont {Steinberg}}, \bibinfo {author} {\bibfnamefont {J.}~\bibnamefont {Tollaksen}},\ and\ \bibinfo {author} {\bibfnamefont {A.~N.}\ \bibnamefont {Jordan}},\ }\bibfield  {title} {\bibinfo {title} {Weak-value amplification and optimal parameter estimation in the presence of correlated noise},\ }\bibfield  {journal} {\bibinfo  {journal} {Physical Review A}\ }\textbf {\bibinfo {volume} {96}},\ \href {https://doi.org/10.1103/physreva.96.052128} {10.1103/physreva.96.052128} (\bibinfo {year} {2017})\BibitemShut {NoStop}%
\bibitem [{\citenamefont {Jordan}\ \emph {et~al.}(2014)\citenamefont {Jordan}, \citenamefont {Martínez-Rincón},\ and\ \citenamefont {Howell}}]{wvatechnoise}%
  \BibitemOpen
  \bibfield  {author} {\bibinfo {author} {\bibfnamefont {A.~N.}\ \bibnamefont {Jordan}}, \bibinfo {author} {\bibfnamefont {J.}~\bibnamefont {Martínez-Rincón}},\ and\ \bibinfo {author} {\bibfnamefont {J.~C.}\ \bibnamefont {Howell}},\ }\bibfield  {title} {\bibinfo {title} {Technical advantages for weak-value amplification: When less is more},\ }\bibfield  {journal} {\bibinfo  {journal} {Physical Review X}\ }\textbf {\bibinfo {volume} {4}},\ \href {https://doi.org/10.1103/physrevx.4.011031} {10.1103/physrevx.4.011031} (\bibinfo {year} {2014})\BibitemShut {NoStop}%
\bibitem [{\citenamefont {Pang}\ \emph {et~al.}(2016)\citenamefont {Pang}, \citenamefont {Alonso}, \citenamefont {Brun},\ and\ \citenamefont {Jordan}}]{pangnoise}%
  \BibitemOpen
  \bibfield  {author} {\bibinfo {author} {\bibfnamefont {S.}~\bibnamefont {Pang}}, \bibinfo {author} {\bibfnamefont {J.~R.~G.}\ \bibnamefont {Alonso}}, \bibinfo {author} {\bibfnamefont {T.~A.}\ \bibnamefont {Brun}},\ and\ \bibinfo {author} {\bibfnamefont {A.~N.}\ \bibnamefont {Jordan}},\ }\bibfield  {title} {\bibinfo {title} {Protecting weak measurements against systematic errors},\ }\bibfield  {journal} {\bibinfo  {journal} {Physical Review A}\ }\textbf {\bibinfo {volume} {94}},\ \href {https://doi.org/10.1103/physreva.94.012329} {10.1103/physreva.94.012329} (\bibinfo {year} {2016})\BibitemShut {NoStop}%
\bibitem [{\citenamefont {Harris}\ \emph {et~al.}(2017)\citenamefont {Harris}, \citenamefont {Boyd},\ and\ \citenamefont {Lundeen}}]{wvasaturation}%
  \BibitemOpen
  \bibfield  {author} {\bibinfo {author} {\bibfnamefont {J.}~\bibnamefont {Harris}}, \bibinfo {author} {\bibfnamefont {R.~W.}\ \bibnamefont {Boyd}},\ and\ \bibinfo {author} {\bibfnamefont {J.~S.}\ \bibnamefont {Lundeen}},\ }\bibfield  {title} {\bibinfo {title} {Weak value amplification can outperform conventional measurement in the presence of detector saturation},\ }\bibfield  {journal} {\bibinfo  {journal} {Physical Review Letters}\ }\textbf {\bibinfo {volume} {118}},\ \href {https://doi.org/10.1103/physrevlett.118.070802} {10.1103/physrevlett.118.070802} (\bibinfo {year} {2017})\BibitemShut {NoStop}%
\bibitem [{\citenamefont {Ueda}\ and\ \citenamefont {Kitahara}(2021)}]{edm}%
  \BibitemOpen
  \bibfield  {author} {\bibinfo {author} {\bibfnamefont {D.}~\bibnamefont {Ueda}}\ and\ \bibinfo {author} {\bibfnamefont {T.}~\bibnamefont {Kitahara}},\ }\bibfield  {title} {\bibinfo {title} {Novel approach to neutron electric dipole moment search using weak measurement},\ }\href {https://doi.org/10.1088/1361-6455/abc5a0} {\bibfield  {journal} {\bibinfo  {journal} {Journal of Physics B: Atomic, Molecular and Optical Physics}\ }\textbf {\bibinfo {volume} {54}},\ \bibinfo {pages} {085502} (\bibinfo {year} {2021})}\BibitemShut {NoStop}%
\bibitem [{\citenamefont {Hu}\ and\ \citenamefont {Zhang}(2017)}]{ligo}%
  \BibitemOpen
  \bibfield  {author} {\bibinfo {author} {\bibfnamefont {M.-J.}\ \bibnamefont {Hu}}\ and\ \bibinfo {author} {\bibfnamefont {Y.-S.}\ \bibnamefont {Zhang}},\ }\href {https://doi.org/10.48550/ARXIV.1707.00886} {\bibinfo {title} {Gravitational wave detection via weak measurements amplification}} (\bibinfo {year} {2017})\BibitemShut {NoStop}%
\bibitem [{\citenamefont {Steck}(2003)}]{steck}%
  \BibitemOpen
  \bibfield  {author} {\bibinfo {author} {\bibfnamefont {D.~A.}\ \bibnamefont {Steck}},\ }\href {https://steck.us/alkalidata/rubidium87numbers.1.6.pdf} {\bibinfo {title} {Rubidium 87 d line data}} (\bibinfo {year} {2003})\BibitemShut {NoStop}%
\end{thebibliography}%
\endgroup
\end{small}

% \printbibliography

\end{document}